\documentclass[fleqn,twoside]{article}
\usepackage{Gc}

\usepackage{amssymb}
\usepackage{amsfonts}

\heads{W. God{\l}owski, M. Szyd{\l}owski and Z.-H. Zhu}
      {Bouncing cosmology caused by Casimir effect}

\begin{document}
\twocolumn[
\Arthead{11}{2005}{4 (44)}{1}{10}

\Title{Constraining bouncing cosmology caused by      \yy
       Casimir effect}

   \Authors{W{\l}odzimierz God{\l}owski\foom 1           
           Marek Szyd{\l}owski\foom 2      
	         }   
           {and Zong-Hong Zhu\foom 3      
	         }   
           {Astronomical Observatory, Jagiellonian University,
 		Orla 171, 30-244 Krak{\'o}w, Poland  (WG and MS)   \\   
           Marc Kac Complex Systems Research Centre, Jagiellonian University,
		Reymonta 4, 30-059 Krak{\'o}w, Poland (MS)}
           {Department of Astronomy, Beijing Normal University,
		Beijing 100875, China}      

\Abstract{
We constrain the Friedman Robertson Walker (FRW) model with
``radiation-like'' contribution to the Friedmann equation against the
astronomical data. We analyze the observational limitations on a
$(1+z)^4$ term from supernovae type Ia (SNIa) data, Fanaroff-Riley
type IIb (FRIIb) radio galaxy (RG) data, baryon oscillation peak and
cosmic microwave background radiation (CMBR) observations. We argue that
it is not possible to determine the energy densities of individual
components scaling like radiation from a kinematic astronomical test.
The bounds for density parameter for total radiation-like term can be
obtained. We find different interpretations of the presence of
scaling like radiation term:
the FRW universe filled with a massless scalar field in a quantum regime
(the Casimir effect), the FRW model in a semi-classical
approximation of loop quantum gravity, the FRW model in the Randall Sundrum
scenario with dark radiation or cosmological model with global rotation.
In this paper we mainly concentrate on the Casimir effect arising from quantum
effects of the scalar field. This contribution can describe decaying part of
cosmological constant. We discuss the back reaction of gravity
on Casimir-type force which is a manifestation of the vacuum fluctuations of
the quantum scalar field at low temperature. It is shown that while the Casimir
energy gives
rise to the accelerating Universe, the cosmological constant term is still
required. We argue that a small negative contribution of a radiation-like term
can reconcile the tension between the observed primordial ${}^4He$ and $D$
abundance. Moreover the presence of such contribution can also remove the
disagreement between the Hubble parameter $H_0$ values obtained from both SNIa
and Wilkinson Microwave Anisotropy Probe (WMAP) satellite data.
}

\RAbstract               
    {Title in Russian}
    {Author(s) in Russian}
    {Text of abstract in Russian}


]  
\email 1 {godlows@oa.uj.edu.pl}
\email 2 {uoszydlo@cyf-kr.edu.pl}
\email 2 {zhuzh@bnu.edu.cn}

\section{Introduction}

From the recent measurements of distant supernovae type Ia
\cite{Riess:2004nr,Astier:2005} we deduce that
the universe is in an accelerating phase of expansion. The effective and
simple explanation of this current state of the Universe requires that
one third of total energy in the universe is non-relativistic dust (dark
matter) and two third is a constituent of negative pressure (dark energy).
While cosmological constant remains
the simplest explanation of
observations of distant supernovae, there appeared serious problem in this
context like: why is the vacuum energy so much smaller than
we except from effective
quantum field calculations? \cite{Weinberg89}.
To better understand how vacuum energy contributes to
the cosmological constant,  we consider the Casimir effect which is a purely
quantum field theory phenomenon. We discuss the back reaction of gravity
on Casimir-type force which is a manifestation of the vacuum fluctuations of
the quantum scalar field at finite temperature. It is shown that while the
Casimir energy gives rise
to accelerating Universe, the cosmological constant is still
required. The FRW model with a Casimir type force contains a term which
scales like negative radiation ($(-)(1+z)^4$).

There are different interpretations of the presence of such a term:
cosmological model with global rotation, Friedmann-Robertson-Walker (FRW)
model in the Randall Sundrum scenario with dark radiation,
FRW universe filled with a massless scalar field in a quantum regime
(Casimir effect), or FRW model in  a semi-classical approximation of loop
quantum gravity.
To constrain the ``radiation-like'' contribution to the Friedmann equation,
we use a variety of astronomical observations, such as SNIa data
\cite{Riess:2004nr,Astier:2005}, FRIIb RG data \cite{Daly04}, baryon oscillation
peak and CMBR observations. Although the obtained bounds on total
density parameters strongly limit the presence of this term in the Friedmann
equation, this does not mean that it is not present, as it is impossible
to determine the energy density of separate components.
We note that the CMBR and big-bang nucleosynthesis (BBN) offer stringent
conditions on this term, which can be regarded as an established upper limit
on any individual components of negative energy density, and therefore
on the Casimir effect, global rotation, discreteness of space following loop
quantum gravity or brane dark radiation.
The plan of the paper is as follows. In the next section, we summarize different
interpretations to the possibility radiation like term in Friedmann equation.
We provide observational constraints on the radiation like term in Section III.
Finally we conclude our work in Section IV.

\section{Different interpretations of the presence of negative radiation
like term in the $H^2(z)$ relation.}

\subsection{Casimir effect}
 
Ishak \cite{Ishak05} distinguished old and new cosmological problems.
While the old problem is related to the order of magnitude of cosmological
constant the new problem is related to this magnitude being of the same
order as the matter density during the present epoch. He pointed out the
relevance of experiments focusing on the Casimir effect in gravitational
and cosmological context \cite{Ishak05}. He argued that while the geometrical
cosmological constant has no quantum properties, the vacuum energy has both
gravitational and quantum properties. In this context the Casimir effect
\cite{Casimir48} seems to be relevant because it can tell us how vacuum energy
contributes to the cosmological constant \cite{Padmanabhan03,Carrol01}.

The Casimir effect can be derived from the general principles of quantum
theory of electromagnetic field \cite{Casimir48} for two uncharged, perfectly
conducting plates in vacuum. They should attract each other with
the force $F \propto d^{-4}$, where $d$ is distance between the plates. This
prediction was verified experimentally and yielded, for electric bodies,
reasonable agreement to the theory \cite{Nesterenko05}.
The Casimir effect is a simple observational consequence of the existence
of quantum fluctuations \cite{Bordag01}. The Casimir force between conducting
plates leads to a repulsive force, like the positive cosmological constant.
It is worthy to mention that Casimir type of contribution arising from the
tachyon condensation is possible \cite{McInnes06}.

Moreover, different laboratory experiments were designed to measure the
Casimir effect with increased precision and then strengthen the constraints
on correction to Newtonian gravitational law \cite{Fischback98}. Therefore, 
the measurement of the thermal Casimir force is promising to obtain stronger 
constraints on non-Newtonian gravity. Although the Casimir force
is very weak (for typical values of $d=0.5 \mu m$ and area of $1$ cm${}^2$, 
the Casimir force is $\simeq 0.2$ dyn) it becomes measurable with a high 
degree of precision \cite{Bressi02}.

For a survey of recently obtained results in the Casimir energy studies see
\cite{Nesterenko05}. The Casimir effect can be regarded as a manifestation
of the quantum fluctuations on the geometry and the topology of the system
boundaries although the Casimir force can be also present with the system
with no boundaries and a compact topology. Therefore, if our Universe has
nontrivial topology then every quantum fields will generate a Casimir-type
force which many authors study on the FRW background space-time. If we consider
a massless scalar field, conformally coupled to gravity, on the background
of the static Einstein universe then its Casimir energy has been shown to
have the form $\alpha/a^4$ with the value of $\alpha=1/(480\pi^2)$
\cite{Ford76}. While for all such fields $\alpha$ is positive because such
models obey the strong energy condition, the computation of Casimir energy
in the cosmological context leads to the conclusion that the Casimir energy
of the scalar fields could drive the inflation in the flat universe with
toroidal topology \cite{Zeldovich84}. It has been recently demonstrated
that there exists a family of quantum scalar fields which give rise to a
repulsive Casimir force in a closed universe \cite{Herdeiro06}.
They ``produce'' Casimir energy scaling like radiation ($\alpha<0$), and
violating the strong energy condition. However, we must remember that all
calculations of the Casimir effect are performed under the assumption of a
quasi adiabatic approximation and their generalization to the case of the
non-static FRW models is extremely difficult.

The finite temperature quantum effects of massless scalar fields on the
background of spacetimes of cosmological models have been considered by many
authors for many years in the context of the Kaluza-Klein theories and 
dynamical reduction of extra dimension process. The main result of these 
investigations was a universal quantum correction in low and high 
temperatures. Assuming the so called adiabatic approximation it can be shown 
that while at low temperature the Casimir energy is always negative and 
proportional to $\alpha/a^{4+d}$ where $d$ is the number of extra dimensions, 
and is positive and proportional to $\alpha/a^{4+d}$ at high temperatures 
\cite{Szydlowski87}. Therefore if $d=0$ we can recover results for the 
standard cosmology which indicate that quantum effects in cosmological models 
at finite temperatures have a universal asymptotic in both high and low 
temperatures. The Casimir energy at low temperatures scales like radiation but 
is still negative. Finally the Casimir effect is significant when the topology 
of the Universe is not trivial \cite{Lachieze-Rey95}. Also Casimir type energy 
can be produced from some extra dimensions \cite{Mazur04,Mazur06}. Recently the
relevance of the Casimir effect in the context of dark energy problem has been
pointed out \cite{Ishak05,Bean,Volovik06,Altaie03,Saharian06}.

When the field occupies some bounded region of the configuration space
then its spectrum is discrete and general vacuum energy $E_0$ depends on
eigenfrequencies $\omega_n$ which can be determined from the geometry of boundary
$\partial \mathcal{M}$; $E_0(\partial \mathcal{M})=\frac{1}{2} \sum \omega_n$.
While the total zero-point energy of the vacuum is infinite in the presence of
boundaries it is modified and $E_0=E_0(\partial \mathcal{M})-E_{0}(0)$
In the case considered by Casimir, when electromagnetic field is confined
between two parallel conducting plates the contribution of unbounded Minkowski
space $-E_0(0)$ should be subtracted. The generally to obtain the physical
value of the vacuum energy we must remember that in quantum field theory
(i.e., in the case of infinite number of degrees of freedom) the observable
quantity is not zero point energy it self, but only its excess, caused
by boundaries \cite{Nesterenko05}.

Recently, the idea that the Casimir effect is responsible for UV ultraviolet
cut-off that renders the total vacuum energy finite was rule out
\cite{Mahajan06}. Therefore the Casimir effect cannot be treated as a natural
cut off leading to the observed cosmological constant value.

The cosmological significance of the Casimir effect was pointed out in many contexts
since the pioneering paper of Zeldovich and Starobinski \cite{Zeldovich84}.
Also, the Casimir effect has been used as an effective mechanism of
compactification of extra dimensions in the Kaluza-Klein cosmological models ---
so called dynamical reduction mechanism \cite{Szydlowski87,Szydlowski88a}.

It was demonstrated that nontrivial topology of a physical space as well as an 
internal space can lead to compactification of an extra dimension
\cite{Szydlowski88b,Szydlowski89}, (for contemporary context see also
\cite{Mazur04}. In these investigations the quantum field theory at finite
temperature is used in calculation of quantum effects and back-reaction
arising from massless scalar fields. It is assumed that the metric of background
space time (FRW space time) is static and conditions for thermodynamical
equilibrium of matter fields are satisfied due to their interactions with
the thermal bath. As a result we obtain universal approximation of quantum
distribution function for massless scalar bosons at high and low temperatures.
The assumption of the quasi-static approximation (characteristic time of
quantum process is much smaller than the characteristic time of cosmic
evolution) enables us to determine the thermodynamical characteristic.
As a result we obtain at low temperatures: $p ( \mathcal{M}^3) = (-) \rho$,
$\rho=(-)\frac{|\alpha|}{a^{D+4}}$, $p'(\mathcal{M}^D)=\frac{4}{D}\rho$,
where $p$ and $p'$ are pressure on physical ($M^3$) and internal ($M^D$)
spaces. $D$ is the dimension of space with additional dimensions, $a$ is the scale
factor.

Because in this case the energy momentum tensor is traceless
(massless scalar field) we define the energy momentum tensor as
$T^{\mu}_{\nu}=\mathrm{diag}(\rho,-p,-p,-p,-p',-p',-p')$. In the standard
space of topology $\mathcal{R} \times \mathcal{M}^3$, where
$\mathcal{M}^3$ is homogeneous and isotropic space with Robertson-Walker
metric we know that quantum effects of scalar field are equivalent to the
effect of fluid with pressure $p=\frac{1}{3}\rho$ and
$\rho =(-)\frac{|\alpha|}{a^4}$. It is a universal approximation effect
of quantum fields originating from massless scalar fields at low
temperatures (Casimir effect). Analogical result was recently obtained by
Herdeiro and Sampaio \cite{Herdeiro06}. The back-reaction problem is considered
by taking the semi-classical equation with $\Lambda$
\begin{equation}
\label{eq:b1}
G_{\mu \nu}+\Lambda g_{\mu \nu}=T^{\mathrm{matter}}_{\mu \nu}+<T^{\phi}_{\mu \nu}>.
\end{equation}
Considering R-W symmetry and matter content in the form of perfect fluid
with energy density $\rho$ and pressure $p$ we obtain
\begin{equation}
\label{eq:b2}
\dot{a}^2+k=\frac{\rho a^2}{3},
\end{equation}
\begin{equation}
\label{eq:b3}
\ddot{a}=-\frac{1}{6}(\rho+3p)a,
\end{equation}
where dot means differentiation with respect the cosmological time and
$\rho$ is the effective energy density
\begin{equation}
\label{eq:b4}
\rho_{\mathrm{eff}}=\Lambda+\frac{\alpha}{a^4},
\end{equation}
where $\alpha$ is the constant of the Casimir force scaling like radiation,
positive for conformally coupled scalar field
in the cosmological context as it was verified long time ago by
Zeldovich and Starobinski \cite{Zeldovich84}. Casimir energy is negative in
a flat universe with toroidal topology. In this case ($\alpha<0$) it is
produced accelerating phase of expansion of the universe because the strong
energy condition $\rho_{\mathrm{eff}}+3p_{\mathrm{eff}}>0$ is violated.
\begin{equation}
\label{eq:b5}
\rho_{\mathrm{eff}}=\Lambda-\frac{|\alpha|}{a^4}+\frac{\rho_{\mathrm{m},0}}{a^3}, \qquad
p_{\mathrm{eff}}=-\Lambda-\frac{|\alpha|}{3a^4}.
\end{equation}

For our aims it is important that the effects of Casimir energy with a negative
value of $\alpha$ which scales like radiation can contribute into the $H^2(z)$
relation---crucial for any kinematic test. It is also interesting that the
same type of contribution to the effective energy density can be produced by
loop quantum theory effects in semi-classical quantum cosmology
\cite{Vandersloot05,Singh05,Hossain05}. These effects give rise to evolutional
scenario in which the initial  singularity is replaced by a bounce.

\subsection{Other interpretations of the Casimir-type term}

There are  many different interpretations of the term in the
Friedmann equation which diminishes with the cosmic scale factor like $a^{-4}$.
The first interpretation comes from the generalized Friedmann equation
on the brane in the Randall and Sundrum scenario \cite{Randall99a,Randall99b}
In the brane world scenario our universe is some sub-manifold
which is embedded in a higher-dimensional space-time called bulk spaces.
While the physical matter fields are confined to this sub-manifold called
brane, the gravity can reside in the higher dimensions. This brane
paradigm was first proposed by Arkani-Hamed et al. \cite{Arkani98} as a means
to reconcile the hierarchy problem between the weak scale and the new Planck
scale $\mathcal{M}_{\mathrm{pl}}$. Randal and Sundrum \cite{Randall99a} solved
the analogous problem between the weak scale and the size of extra dimension
by introducing non-compact extra dimensions. In their model our universe is
represented by a three brane embedded in a 5 dimensional anti de Sitter space.
The cosmological evolution of such brane universes was extensively
investigated by several authors (see for example \cite{Maartens00}).
This way, the Einstein equations restricted to the brane reduce to some
generalization of the FRW equation. Two additional terms contribute into
the $H^2$ relation  \cite{Vishwakarma03}. The $\rho^2$ term arises from the
imposition of a junction condition for the scale factor on the surface
on the brane. This term decays rapidly as $a^{-6}$ for dust or as $a^{-8}$
for the radiation dominated early universe. This term should be significant
only in the very early universe \cite{Vishwakarma03,Godlowski04}.

The second term is of considerable interest for us because it scales like
radiation with a negative constant $\alpha$. Hence it is called dark radiation.
This term arises from the non-vanishing electric part of the five dimensional
Weyl tensor.
Mathematically both negative and positive values of $\alpha$ are possible.

Dark radiation strongly effects both BBN and CMBR.
It was demonstrated by Ichiki et al. \cite{Ichiki03} that BBN limits the
possible contribution from dark radiation just before $e^{+}e^{-}$ annihilation
epoch. They gave limits on the possible contribution  of dark radiation as
$-1.23<\rho_{dr}/\rho_{\gamma} \le 0.11$ from BBN and
$-0.41<\rho_{dr}/\rho_{\gamma} \le 0.105$ at the 95\% confidence level
from CMBR measurements. Let us note that small negative contribution of dark
radiation can also reconcile the tension between the observed
${}^4\mathrm{He}$ and $D$ abundances \cite{Ichiki02}.

Another interpretation of the presence of the negative radiation like term is rotation
in the Newtonian cosmology. When we consider Newtonian cosmology following
Senovilla et al. \cite{Senovilla98} then we can define, homogeneous Newtonian
cosmology as $\rho$ and $p$ having no spatial dependence i.e $\rho=\rho(t)$
and $p=p(t)$ while  we  assume that the velocity vector fields depends linearly
on the spatial coordinates.
In such a case we obtain  equation which represents shear-free
Newtonian cosmologies with expansion and rotation which is well known as the
Heckmann-Sch{\"u}cking model \cite{Heckmann59}.
\begin{equation}
\label{eq:5}
\dot{a}^2=\frac{\rho(t_0)}{3a}-\frac{2\omega^2}{3a^2}+C
\end{equation}
where $C$ is an arbitrary constant. We interpret it in terms of curvature
constant although in the Newtonian spacetime the curvature is zero.
For our aims it is important that the effect of rotation produce negative
term scaling like $(1+z)^4$ in the Newtonian analogue of the Friedmann equation.

In the Newtonian cosmology in contrast to general relativity effect of rotation
are not necessary related to non-vanishing shear. The homogeneous universe
with non-vanishing shear basing on general relativity may expand and rotate
relative to local gyroscopes. The problem of relation between the rotation of
the universe and origin of the rotation of galaxies was investigated in
\cite{Li98} and \cite{Godlowski03a,Godlowski05,Aryal06}.
Also, the role of rotation of objects in the Universe, their significance
and astronomical measurements was recently addressed by
\cite{Vishwakarma04,Godlowski03b}.

To compare the results of analyses for general
relativistic model with results obtained further in the paper
we formally consider $\Omega_{k,0}\ne 0$, although the satisfactory
interpretation of curvature density parameter $\Omega_{k,0}$ can be
found in general relativity (see also \cite{Godlowski03b}).

\EFigure{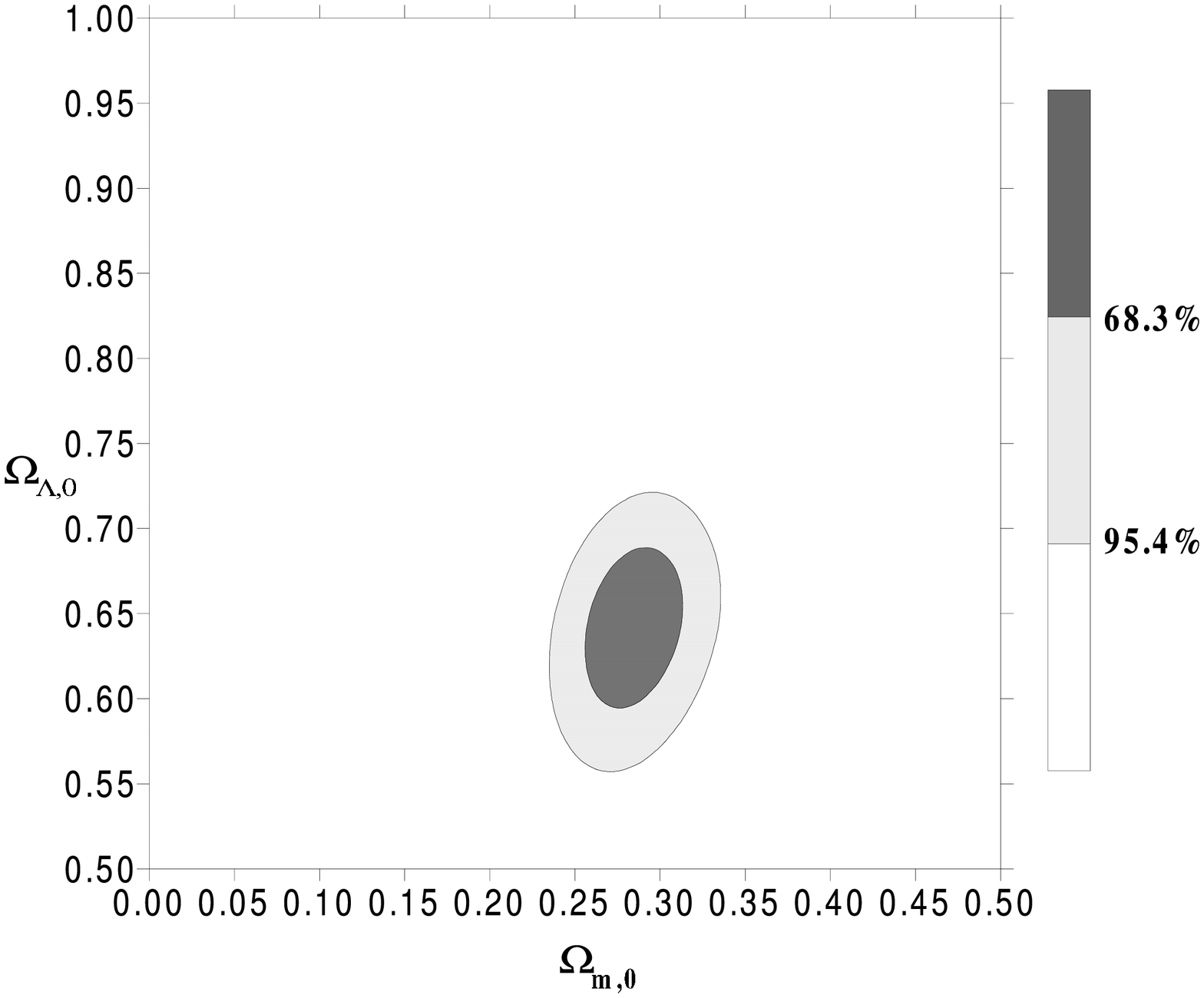}
{The $68.3\%$ and $95.4\%$ confidence levels (obtained from combined
analysis of SN+RG+SDSS+CMBR) on the ($\Omega_{\mathrm{m},0},\Omega_{\Lambda,0}$) 
plane. \label{fig:1}}

\EFigure{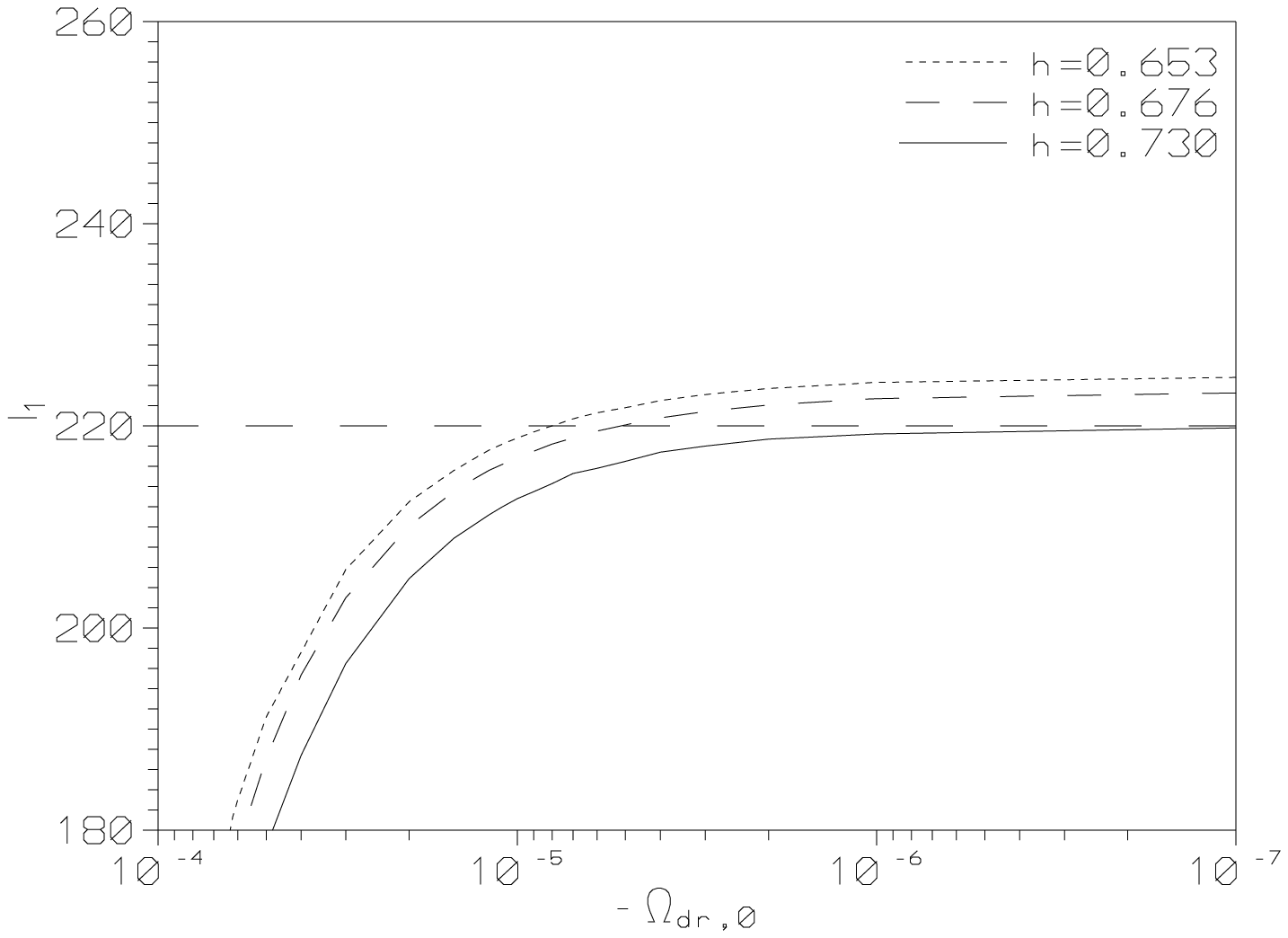}{The location of the first peak
$l_1$ as a function of
$-\Omega_{dr,0}$. Note that $l_1 \simeq 220$ for $h=0.73$ favour
$\Omega_{dr,0} \simeq 0$, while for $h=0.65$ and $h=0.67.6$ lead to
$\Omega_{dr,0} \ne 0$.
\label{fig:2}}

\section{Observational constraints on the FRW model parameter with the radiation-like term}

Cosmological model are usually tested against observations. One of the most
popular test is based on the luminosity distance $d_L$ of the supernovae Ia
as a function of redshift \cite{Riess98}. However for the distant SNIa,
one can not directly observe their luminosity distances $d_L$ but their
apparent magnitudes $m$ and redshifts $z$. The absolute magnitude $\mathcal{M}$ 
of the supernovae is related to its absolute luminosity $L$. Since we could 
obtain the following relation between distance modulus $\mu$,
the luminosity distance $d_L$, the observed magnitude $m$ and the absolute
magnitude $M$:
\begin{equation}
\label{eq:11}
\mu \equiv  m - M = 5\log_{10}d_{L} + 25=5\log_{10}D_{L} + \mathcal{M}
\end{equation}
where $D_{L}=H_{0}d_{L}$ and $\mathcal{M} = - 5\log_{10}H_{0} + 25$.
We could compute the luminosity distance of a supernova as the function of
redshift:
\begin{equation}
\label{eq:12}
d_L(z) =  (1+z) \frac{c}{H_0} \frac{1}{\sqrt{|\Omega_{k,0}|}} \mathcal{F}
\left( H_0 \sqrt{|\Omega_{k,0}|} \int_0^z \frac{d z'}{H(z')} \right)
\end{equation}
where
\begin{eqnarray}
\left(\frac{H}{H_0}\right)^2 &=& \Omega_{\mathrm{m},0}(1+z)^{3}+\Omega_{k,0}(1+z)^{2} 
+\Omega_{\mathrm{r},0}(1+z)^{4} \nonumber \\
&&+\Omega_{\mathrm{dr},0}(1+z)^{4}+\Omega_{\Lambda,0},
\label{eq:12a}
\end{eqnarray}
$\Omega_{k,0} = - \frac{k}{H_0^2}$ and
$\mathcal{F} (x) \equiv (\sinh (x), x,\sin (x))$ for $k<0, k=0, k>0$,
respectively. We assumed $\Omega_{\mathrm{r},0} = \Omega_{\gamma,0} + \Omega_{\nu,0}
= 2.48 h^{-2} \times 10^{-5} + 1.7 h^{-2} \times 10^{-5}\simeq 0.0001$
\cite{Vishwakarma03}.

Daly and Djorgovski \cite{Daly03} suggested to include in the analysis not only 
supernovae but also radio galaxies (see also \cite{Zhu04,Puetzfeld05,Godlowski06}).
In such a case, it is useful to use the coordinate distance defined as
\begin{equation}
\label{eq:12b}
y(z)=\frac{H_0 d_L(z)}{c(1+z)}.
\end{equation}

Daly and Djorgovski \cite{Daly04} have compiled a sample comprising the data
on $y(z)$ for 157 SNIa in the Riess et al. Gold dataset \cite{Riess:2004nr} and
20 FRIIb radio galaxies. In our data sets we also include 115 SNIa compiled
by Astier et~al. \cite{Astier:2005}.

It is clear from eq. (\ref{eq:12a}) and (\ref{eq:12b}) that the coordinate 
distance does not depend on the value of $H_0$. Unfortunately we do not know
coordinate distance $y(z)$ for supernovae. This distance must be computed
from the luminosity distance (or the distance modulus $\mu$) and for such
a computation a knowledge of the value of $H_0$ is required.
For both supernovae sample we choose the values of $H_0$ which were used in the original papers.
We used the distance modulus presented in Ref.~\cite{Riess:2004nr,Astier:2005} for
the calculation of the coordinate distance. For each sample we choose the values
of $H_0$ appropriate to the data sets. For Riess et al.'s Gold sample
we fit the value of $h=0.646$ as the best fitted value and this value is used
for calculation of coordinate distance for SNIa belonging to this sample.
In turn the value $h=0.70$ was assumed in the calculations
of the coordinate distance for SNIa belonging to Astier et~al.'s sample,
because the distance moduli $\mu$ presented in Ref.~\cite[Tab.~8]{Astier:2005}
was calculated with such an arbitrary value of $h=0.70$.
The error of the coordinate distance can be computed as
\begin{eqnarray}
\sigma^2(y_i) &=& \left(\frac{10^{\frac{\mu_i}{5}}}{c \left(1+z \right) 10^5} \right)^2
\times \nonumber \\
&& \times \left(\sigma^2(H_0)+\left(\frac{H_0 \ln{10}}{5}\right)^2\sigma^2(\mu_i)\right)
\label{eq:12c}
\end{eqnarray}
where $\sigma_i(\mu_i)$ denotes the statistical error of distance modulus
determination (note that for Astier et~al.'s sample the intrinsic dispersion was
also included) and $\sigma(H_0)= 0.8$ km/s Mpc denotes error in $H_0$ measurements.

We included to our constraints obtain from extragalactic analysis:
the baryon oscillation peaks (BOP)
detected in the Sloan Digital Sky Survey (SDSS)
\cite{Eisenstein:2005}. They found that value of $A$
\begin{equation}
\label{eq:16}
A \equiv \frac{\sqrt{\Omega_{\mathrm{m},0}}}{E(z_1)^{\frac{1}{3}}}
\left(\frac{1}{z_1\sqrt{|\Omega_{k,0}|}}
\mathcal{F} \left( \sqrt{|\Omega_{k,0}|} \int_0^{z_1} \frac{d z}{E(z)} \right)
\right)^{\frac{2}{3}}
\end{equation}
(where $E(z) \equiv H(z)/H_0$ and $z_1=0.35$) is equal $A=0.469 \pm 0.017$.
The quoted uncertainty corresponds to one standard deviation, where a
Gaussian probability distribution has been assumed.

Another constraint which we also include in our analysis is the
so called the (CMBR) ``shift parameter''
\begin{equation}
\label{eq:17}
R \equiv \sqrt{\Omega_{\mathrm{m},0}} \, y(z_{lss})=
\sqrt{\frac{\Omega_{\mathrm{m},0}}{|\Omega_{k,0}|}}
\mathcal{F} \left(\sqrt{|\Omega_{k,0}|} \int_0^{z_{lss}} \frac{d z}{E(z)}\right)
\nonumber
\end{equation}
where $R_0=1.716 \pm 0.062$ \cite{Wang04}.

In our combined analysis, we can obtain a best fit model by minimizing the
pseudo-$\chi^2$ merit function \cite{Cardone05}
\begin{eqnarray}
\chi^{2} &=& \chi_{\mathrm{SN+RG}}^{2}+\chi_{\mathrm{SDSS}}^{2} +\chi_{\mathrm{CMBR}}^{2}
\nonumber \\
&=& \sum_{i}\left(\frac{y_{i}^{\mathrm{obs}}-y_{i}^{\mathrm{th}}}{\sigma_{i}(y_{i})}\right)^{2}+
\left(\frac{A^{\mathrm{mod}}-0.469}{0.017}\right)^{2} \nonumber \\
&& + \left(\frac{R^{\mathrm{mod}}-1.716}{0.062}\right)^{2},
\label{eq:18}
\end{eqnarray}
where $ A^{\mathrm{mod}}$ and $R^{\mathrm{mod}}$ denote the values of $A$ and $R$
obtained for a particular set of the model parameter.
For Astier et al. SNIa \cite{Astier:2005} sample additional error in $z$
measurements were taken into account. Here $\sigma_{i}(y_{i})$ denotes
the statistical error (including error in $z$ measurements) of the coordinate
distance determination.

Constraints for the cosmological parameters, can be obtain by minimizing
the following likelihood function $\mathcal{L}\propto \exp(-\chi^{2}/2)$.
One should note that when we are interested in constraining a particular model
parameter, the likelihood function marginalized over the remaining parameters
of the model should be considered \cite{Cardone05}.
This method was used in the case of negative $(1+z)^4$ type
contribution only in the paper \cite{Godlowski06}. In the present paper
we avoid from constrain that a priori the $(1+z)^4$ term must be negative.

Our results are presented in Table~\ref{tab:1}, Table~\ref{tab:2}
and Fig.~\ref{fig:1}. Table~\ref{tab:1} refers to the minimum $\chi^2$ method,
whereas Table~\ref{tab:3} shows the results from the marginalized likelihood
analysis.

From our combined analysis (SN+RG+SDSS+CMBR) we obtain as the best fit
a flat (or nearly flat universe) with $\Omega_{\mathrm{m},0} \simeq 0.3$,  and
$\Omega_{\Lambda,0} \simeq 0.7$. For the dark radiation term, we obtain the
stringent bound
$\Omega_{\mathrm{Totr},0}=\Omega_{\mathrm{r},0}+\Omega_{\mathrm{dr},0} \in (-0.00017, 0.00857)$
at the 95\% confidence level. It lead for limit on dark radiation
$\Omega_{\mathrm{dr},0} \in (-0.00027, 0.0847)$
($\Omega_{\mathrm{Totr},0} \in (-0.00017, 0.0857)$. This results mean that
the positive value of dark radiation term is preferred
($\Omega_{\mathrm{dr},0}>0$), however small negative contribution of dark
radiation is also available. One should note that when we used only SN and RG
data only, we obtain value $\Omega_{\mathrm{m},0}$ close to zero what seems
to be unrealistic. It is the reason that we repeat our analysis with prior
$\Omega_{\mathrm{m},0}=0.3$ \cite{Peebles:2002gy}. In such a case, we again
obtain flat (or nearly flat universe) as a best fit. Contribution of
dark energy is small and positive, but also in this case small negative
contribution of dark radiation is available. From combined analysis we
obtain limit on dark radiation $\Omega_{\mathrm{dr},0} \in (-0.00037, 0.00727)$
($\Omega_{\mathrm{Totr},0} \in (-0.00027, 0.00737)$). Our results shows
that in the present epoch contribution of the dark radiation, if it exist, is small
and gives only small corrections to the $\Lambda$CDM model in the low redshift.

We use the Akaike information criteria (AIC) \cite{Akaike:1974} and the 
Bayesian information criteria (BIC) \cite{Schwarz:1978} to select model 
parameters providing the preferred fit to data. Usually incorporating
new parameters increases the quality of the fit. The question is, if it
increases it significantly enough. The information criteria put a threshold 
which must be exceeded in order to assert an additional parameter to be 
important in explanation of a phenomenon. The discussion how high this threshold
should be caused the appearing of many different criteria. The AIC and BIC 
(for review see \cite{BA2004}) are most popular and used in everyday statistical 
practices. The question is whether the AIC and BIC information criteria of 
model selection provide sufficient arguments for incorporation of new parameters.
The usefulness of using the information criteria of model selection was 
demonstrated by Liddle \cite{Liddle:2004nh} and Parkinson et al. \cite{Parkinson:2005}
Of course,in any case, some future observational data, as well as theoretical
considerations, may give arguments in favour of additional parameter.

The AIC is defined in the following way \cite{Akaike:1974}
\begin{equation} \label{eq:111}
\mathrm{AIC} = - 2\ln{\mathcal{L}} + 2d
\end{equation}
where $\mathcal{L}$ is the maximum likelihood and $d$ is a number of the free model
parameters. The best model with a parameter set providing the preferred fit to
the data is that minimizes the AIC. It is interesting that the AIC also
arises from an approximate minimization of the Kulbak-Leibner information
entropy \cite{Sakamoto86}.

The BIC introduced by Schwarz \cite{Schwarz:1978} is defined as
\begin{equation} \label{eq:112}
\mathrm{BIC} = - 2\ln{\mathcal{L}} + d\ln{N}
\end{equation}
where $N$ is the number of data points used in the fit. Comparing these criteria,
one should note that the AIC tends to favour models with large number of 
parameters unlike the BIC, because the BIC penalizes new parameters more 
strongly. It is the reason that the BIC provides a more useful approximation 
to the full statistical analysis in the case of no priors on the set of model 
parameters \cite{Parkinson:2005}. It makes this criterion especially suitable 
in the context of cosmological applications.

Please note that while the AIC is useful in obtaining upper limit to the
number of parameters which should be incorporated to the model, the BIC is
more conclusive. Of course only the relative value between the BIC of different
models has statistical significance. The difference of $2$ is treated as
a positive evidence (and $6$ as a strong evidence) against the model with the
larger value of the BIC \cite{Jeffreys:1961,Mukherjee:1998wp}.
If we do not find any positive evidence from information criteria the
models are treated as a identical and eventually additional parameters are
treated as not significant. The using of
the BIC seems to be especially suitable whenever the complexity of reference
does not increase with the size of data set.  Liddle \cite{Liddle:2004nh}
noted that in cosmology, a new parameter is usually a quantity set to zero in
a simpler base model and if the likelihood function is a continuous function
of its parameters it will increase as the parameter varies in either the
positive or negative direction. The problem of classification of the
cosmological models on the light of information criteria on the base of the
astronomical data was discussed in our previous papers
\cite{Godlowski05a,Szydlowski06a,Szydlowski06b,Szydlowski06c,Szydlowski06d}.

Our results are presented in Table~\ref{tab:5}. It is clear that in the
light of informative criterion model with dark energy do not increase
fit significantly. It confirm our conclusion that dark energy term,
if it exist, is small in the present epoch.
Using the prior $\Omega_{\mathrm{m},0}=0.3$
\cite{Peebles:2002gy} does not change our conclusion. It shows that using such
a prior in the light of information criteria  is realistic.

Please also note that if
$\Omega_{\mathrm{Totr},0}=\Omega_{\mathrm{r},0}+\Omega_{\mathrm{dr},0}<0$,
then we obtain a bouncing scenario \cite{Molina99,Tippett04,Szydlowski05}
instead of a big bang.
For $\Omega_{\mathrm{m},0}=0.3$, $\Omega_{\mathrm{dr},0}=-0.00027$ and $h=0.65$
bounces ($H^2=0$) appear for $z \simeq 1800$. In this case, the BBN epoch never
occurs and all BBN predictions would be lost.

This results shows that more stronger constraints for model parameters is
required. To obtain such a constraints on the model parameters, it is useful to 
use the CMBR observations. The hotter and colder spots in the CMBR can be interpreted
as acoustic oscillations in the primeval plasma during the last scattering.
In such a case the locations of the peaks in the CMBR power spectrum  are very
sensitive to variations in the model parameters. Therefore, the location of these
peaks can also be used for constraining the parameters of cosmological models.
The acoustic scale $\ell_{A}$ which gives the locations of the peaks is defined as
\begin{equation}
\ell_{A} = \pi \frac{\int_{0}^{z_{\rm dec}} \frac{d z'}{H(z')}}
{\int_{z_{\rm dec}}^{\infty} c_{s} \frac{d z'}{H(z')}}
\end{equation}
where, for the flat model, equation~(\ref{eq:12a}) reduces to
\begin{eqnarray}
H(z) &=& H_{0} \left[ \Omega_{\mathrm{m},0}(1+z)^3 + \Omega_{\mathrm{r},0}(1+z)^4 
\right. \nonumber \\
&& \left. +\Omega_{\mathrm{dr},0}(1+z)^4  +\Omega_{\Lambda,0} \right]^{1/2},
\end{eqnarray}
where $c_{\mathrm{s}}$ is the speed of sound in plasma.
Knowing the acoustic scale we can determine the location of the $m$-th peak
$\ell_{m} \sim \ell_{A}(m- \phi_{m})$ where $\phi_{m}$ is the phase shift
caused by the plasma driving effect. The CMBR temperature angular power
spectrum provides the locations of the first two peaks
$\ell_{1} = 220.1_{-0.8}^{+0.8}$,  $\ell_{2} = 546_{-10}^{+10}$
\cite{Spergel:2003}. Using three years of WMAP data, Spergel et~al. obtained
that the Hubble constant $H_{0}=73$ km/s Mpc, the baryonic matter density
$\Omega_{\mathrm{b},0} = 0.0222h^{-2}$, and the matter density
$\Omega_{\mathrm{m},0} = 0.128h^{-2}$ \cite{Spergel06}, which are in good agreement
with the observation of position of the first peak (see Fig.~\ref{fig:2})
but lead (assuming the $\Lambda$CDM model) to a value $\Omega_{\mathrm{m},0}=0.24$.
It mean that there is disagreement between $H_0$ values obtained from SNIa and
WMAP. We compute the location of the first peak as a function of
$\Omega_{\mathrm{dr},0}$ assuming $H_0=65$ km/s Mpc ($\Omega_{\mathrm{m},0}=0.3$).
Separately we repeat our computation using the latest Riess et al. result
which obtain $\Omega_{\mathrm{m},0}=0.28$ ($H_0$=67.6 km/s Mpc) \cite{Riess06}.

From Fig.~\ref{fig:2} it is easy to see that we can obtain agreement
with the observation of the location of the first peak for non-zero values of
the parameter $\Omega_{\mathrm{dr},0}$ (Fig.~\ref{fig:2}) both for
$\Omega_{\mathrm{m},0}=0.28$ and $\Omega_{\mathrm{m},0}=0.30$. We obtain
$-1.05\times 10^{-5}<\Omega_{\mathrm{dr},0}<-0.5\times 10^{-5}$ at the $95\%$
confidence level for the case $\Omega_{\mathrm{m},0}=0.3$ while
$-0.75 \times 10^{-5}<\Omega_{\mathrm{dr},0}< -0.25 \times 10^{-5}$ at the $95\%$
confidence level for the case $\Omega_{\mathrm{m},0}=0.28$
Please note that our limits are stronger than that obtained by Ichiki et~al.
\cite{Ichiki03}, which provides bounds of
$-7.22 \times 10^{-5}<\Omega_{\mathrm{dr},0} \le 0.65 \times 10^{-5}$
(in the case of the BBN)
and $-2.41 \times 10^{-5}<\Omega_{\mathrm{dr},0} \le 0.62 \times 10^{-5}$
(in the case of the CMBR).

In all  cases the obtained values of $\Omega_{\mathrm{dr},0}$ are in agreement
with the result obtained from the combined analysis because the $2\sigma$ confidence
interval for this parameter obtained from this analysis contains the area allowed
from the CMBR. Most important conclusion is, that while the combined analysis
allowed the possibility that $\Omega_{\mathrm{dr},0}$ is equal to zero, the CMBR
location of the first peak seems to exclude this case both for $h=0.65$ and
$h=0.67$.

\section{Conclusion}

In the paper we analysed the observational constraints on the $(1+z)^4$-type
contribution in the Friedmann equation. Because it the present paper
we mainly concentrate on Casimir effect arising from quantum effects of the
scalar field the constraints for negative $(1+z)^4$-type contribution are
in our special interest. The analysis of SNIa data as well as
both SNIa and FRIIb radio galaxies (with and without priors going from baryon
oscillation peaks and CMBR ``shift parameter'') shows that the values of $\chi^2$
statistics are lower for model with radiation like term, than for the $\Lambda$CDM
model. However, information criteria show that using such a term does not
increase the quality of the fit significantly. BIC even favour $\Lambda$CDM
model over our model of Bouncing Cosmology (with dark radiation). However
this preference is weak. This results show that $(1+z)^4$ term
is not significant in the present epoch of the Universe.

We show that there are several  interpretations of the $(1+z)^4$-type
contribution and we discussed different proposals for
the presence of such a term. Unfortunately, it is not possible, with present
kinematic astronomical test, to determine the energy densities of individual
components  scales like radiation. However we show that some stringent bounds
on the value of this total contribution can be given.
Combined analysis of SNIa data and FRIIb radio galaxies using  baryon
oscillation peaks and CMBR ``shift parameter'' give rise to a concordance
universe model
which is almost flat with $\Omega_{{\rm m},0} \simeq 0.3$. From the
above-mentioned combined analysis, we obtain an constraint for the term which
scales like radiation $\Omega_{\mathrm{Totr},0} \in (-0.00017, 0.00857)$ which
leads to bounds on the dark radiation term
$\Omega_{\mathrm{dr},0} \in (-0.00027, 0.0847)$ at the 95\% confidence level.
 This is a stronger limit than obtained previously by us from
SNIa data only \cite{Godlowski03b}.
Our model with a small contribution of dark radiation type can also resolve
the disagreement between $H_0$ values obtained from SNIa and WMAP.
For $\Omega_{{\rm m},0} =0.3$  ($H_0=65$ km/s Mpc)
we find new stringent limits on the negative component scaling like radiation
from the location of the peak in the CMBR power spectrum,
$-1.05\ 10^{-5}<\Omega_{dr,0}<-0.5\ 10^{-5}$ at the 95\% confidence level.
This bound is stronger than that obtained  from BBN and CMBR by Ichiki et~al.
\cite{Ichiki03}.

In this paper we have especially studied the advance of initial singularity
using back reaction gravity quantum effect at low temperatures (Casimir effect).
Casimir force arising from the quantum effect of massless scalar field
give rise to a $(-)(1+z)^4$ correction whose effect depends upon the geometry
and nontrivial topology of the space. Typically this type of correction
is thought to be important at the late time of evolution of the universe.
We have shown that Casimir effect could  remove initial
singularity which would be replaced by bounce.
However, from the observational limits
we obtain that bounce does not appear which means that a big bang
scenario is strongly favoured instead of bounce. Our limit
$(-)\Omega_{dr,0}<1.05\ 10^{-5}$  obtain from position of the first
peak in the power spectrum of CMBR leads to
$\Omega_{\mathrm{Totr},0}=\Omega_{dr,0}+\Omega_{r,0}>0$. This implies that
$H^2(z)$ is always greater than zero ($H^2(z)>0$) and bounce does not
appear which means that a big bang scenario is strongly favoured instead of
bounce.

\Acknow
{M. Szyd{\l}owski was supported by project ``COCOS'' No. MTKD-CT-2004-517186.
Z.-H. Zhu acknowledges support from the National Natural Science Foundation of
China, under Grant No. 10533010, and SRF, ROCS, SEM of China.
The authors also thank Dr. A. G. Riess, Dr. P. Astier and Dr. R. Daly for
the detailed explanation of their data samples.}

\begin{table*}
\caption{Results of the statistical analysis of the model with
radiation like term obtained from $\chi^2$ best fit.
The upper section of the table represents the constraint $\Omega_{k,0}=0$ (flat model).
\label{tab:1}}
\begin{tabular}{@{}p{4.2cm}rrrrr}
\hline  \hline
sample & $\Omega_{k,0}$ & $\Omega_{\mathrm{m},0}$ & $\Omega_{\mathrm{Totr},0}$ & $\Omega_{\Lambda,0}$ & $\chi^2$ \\
\hline
SN             &   -   &$ 0.01$ & 0.133   & 0.807 & 295.9     \\
SN+RG          &   -   &$ 0.11$ & 0.123   & 0.767 & 319.5     \\
SN+RG+SDSS     &   -   &$ 0.28$ & 0.022   & 0.698 & 320.1     \\
SN+RG+SDSS+CMBR&   -   &$ 0.30$ & 0.00054 & 0.699 & 322.3     \\
\hline
SN             &  0.01 &$ 0.00$ & 0.186   & 0.804 & 295.9     \\
SN+RG          &  0.08 &$ 0.00$ & 0.169   & 0.751 & 319.5     \\
SN+RG+SDSS     & -0.11 &$ 0.28$ & 0.049   & 0.781 & 319.5     \\
SN+RG+SDSS+CMBR&  0.03 &$ 0.29$ & 0.00270 & 0.657 & 320.9     \\
\hline
\end{tabular}
\end{table*}

\begin{table*}
\caption{Results of the statistical analysis of the model with
radiation-like term. The values of the model parameters are obtained from
marginalized likelihood analysis. We present maximum likelihood value with
$68.3\%$ confidence ranges. The upper section of the table represents
the constraint $\Omega_{k,0}=0$
(flat model). \label{tab:2}}
\begin{tabular}{@{}p{4.2cm}cccc}
\hline  \hline
sample & $\Omega_{k,0}$ & $\Omega_{\mathrm{m},0}$ & $\Omega_{\mathrm{Totr},0}$ & $\Omega_{\Lambda,0}$ \\
\hline
SN             &   -   & $0.00^{+0.23}_{-0.00}$ &$0.164^{+0.045}_{-0.263}$     &$0.78^{+0.03}_{-0.08}$ \\
SN+RG          &   -   & $0.10^{+0.16}_{-0.10}$ &$0.128^{+0.057}_{-0.109}$     &$0.77^{+0.03}_{-0.08}$ \\
SN+RG+SDSS     &   -   & $0.28^{+0.02}_{-0.02}$ &$0.022^{+0.015}_{-0.014}$     &$0.69^{+0.02}_{-0.01}$ \\
SN+RG+SDSS+CMBR&   -   & $0.30^{+0.02}_{-0.01}$ &$0.00056^{0.00162}_{-0.00075}$&$0.69^{+0.01}_{-0.02}$ \\
\hline
SN             &$-0.26^{+0.25}_{-0.28}$ &$ 0.00^{+0.57}_{-0.00}$ &$0.091^{+0.053}_{-0.201}$       &$0.86^{+0.12}_{-0.13}$   \\
SN+RG          &$-0.19^{+0.24}_{-0.27}$ &$ 0.00^{+0.55}_{-0.00}$ &$0.081^{+0.058}_{-0.196}$       &$0.81^{+0.12}_{-0.12}$   \\
SN+RG+SDSS     &$-0.11^{+0.15}_{-0.16}$ &$ 0.28^{+0.02}_{-0.02}$ &$0.051^{+0.042}_{-0.042}$       &$0.78^{+0.12}_{-0.12}$   \\
SN+RG+SDSS+CMBR&$ 0.06^{+0.04}_{-0.04}$ &$ 0.28^{+0.02}_{-0.01}$ &$0.00334^{+0.00267}_{-0.00226}$ &$0.64^{+0.04}_{-0.03}$   \\
\hline
\end{tabular}
\end{table*}

\begin{table*}
\caption{Results of the statistical analysis of the model with the 
radiation like term obtained from $\chi^2$ best fit with the assumption $\Omega_{\mathrm{m},0}=0.3$.
The upper section of the table represents the constraint $\Omega_{k,0}=0$ (flat model).}
\begin{tabular}{@{}p{4.2cm}rrrrr}
\hline  \hline
sample & $\Omega_{k,0}$ & $\Omega_{\mathrm{m},0}$ & $\Omega_{\mathrm{Totr},0}$ & $\Omega_{\Lambda,0}$ & $\chi^2$ \\
\hline
SN             &   -   &$ 0.30$ & $0.009$   & 0.691 & 297.3     \\
SN+RG          &   -   &$ 0.30$ & $0.010$   & 0.690 & 320.2     \\
SN+RG+SDSS     &   -   &$ 0.30$ & $0.012$   & 0.688 & 321.0     \\
SN+RG+SDSS+CMBR&   -   &$ 0.30$ & $0.00054$ & 0.699 & 322.3     \\
\hline
SN             & -0.19 &$ 0.30$ & $0.056$   & 0.834 & 295.9     \\
SN+RG          & -0.13 &$ 0.30$ & $0.042$   & 0.788 & 319.5     \\
SN+RG+SDSS     & -0.09 &$ 0.30$ & $0.034$   & 0.756 & 320.7     \\
SN+RG+SDSS+CMBR&  0.03 &$ 0.30$ & $0.00182$ & 0.668 & 321.4     \\
\hline
\end{tabular}
\label{tab:3}
\end{table*}

\begin{table*}
\caption{Results of the statistical analysis of the model with the 
radiation like term. The values of the model parameters are obtained from the 
marginalized likelihood analysis with the assumption $\Omega_{\mathrm{m},0}=0.3$.
We present maximum likelihood value with $68.3\%$ confidence ranges. The upper 
section of the table represents the constraint $\Omega_{k,0}=0$ (flat model).}
\begin{tabular}{@{}p{4.2cm}cccc}
\hline  \hline
sample & $\Omega_{k,0}$ & $\Omega_{\mathrm{m},0}$ & $\Omega_{\mathrm{Totr},0}$ & $\Omega_{\Lambda,0}$ \\
\hline
SN             &   -   & $0.30$ & $0.009^{+0.011}_{-0.010}$      &$0.69^{+0.01}_{-0.02}$ \\
SN+RG          &   -   & $0.30$ & $0.010^{+0.011}_{-0.010}$      &$0.69^{+0.01}_{-0.02}$ \\
SN+RG+SDSS     &   -   & $0.30$ & $0.012^{+0.010}_{-0.011}$      &$0.68^{+0.02}_{-0.01}$ \\
SN+RG+SDSS+CMBR&   -   & $0.30$ & $0.00054^{+0.00162}_{-0.00070}$&$0.69^{+<0.01}_{-<0.01}$ \\
\hline
SN             &$-0.19^{+0.16}_{-0.17}$ &$ 0.30$ &$0.057^{+0.043}_{-0.072}$      &$0.87^{+0.13}_{-0.12}$   \\
SN+RG          &$-0.13^{+0.15}_{-0.16}$ &$ 0.30$ &$0.042^{+0.040}_{-0.039}$      &$0.82^{+0.12}_{-0.12}$   \\
SN+RG+SDSS     &$-0.09^{+0.15}_{-0.15}$ &$ 0.30$ &$0.034^{+0.039}_{-0.038}$      &$0.61^{+0.12}_{-0.11}$   \\
SN+RG+SDSS+CMBR&$ 0.04^{+0.03}_{-0.03}$ &$ 0.30$ &$0.00217^{+0.00237}_{-0.00173}$&$0.65^{+0.03}_{-0.03}$   \\
\hline
\end{tabular}
\label{tab:4}
\end{table*}

\begin{table*}
\caption{The values of the AIC and BIC for the 
$\Lambda$CDM model and Bouncing Cosmology
model (with dark radiation) without and with prior $\Omega_{\mathrm{m},0}=0.3$.
The upper section of the table represents the constraint $\Omega_{k,0}=0$
(flat model).}
\begin{tabular}{c|cc|cc|cc}
\hline \hline
\multicolumn{1}{c}{}&
\multicolumn{2}{c}{$\Lambda$CDM}&
\multicolumn{2}{c}{BC}&
\multicolumn{2}{c}{BC($\Omega_{\mathrm{m},0}=0.3$)}\\
\hline \hline
sample & AIC & BIC & AIC & BIC & AIC & BIC \\
\hline
SN             & 299.5& 303.1& 299.9 & 307.1 & 299.3 & 302.9  \\
SN+RG          & 322.4& 326.1& 323.5 & 330.9 & 322.2 & 325.9  \\
SN+RG+SDSS     & 324.4& 328.1& 324.1 & 331.5 & 323.0 & 326.7  \\
SN+RG+SDSS+CMBR& 324.5& 328.2& 326.3 & 333.7 & 324.3 & 328.0  \\
\hline
SN             & 300.0& 307.2& 301.9 & 312.7 & 299.9 & 307.1  \\
SN+RG          & 323.5& 330.9& 325.5 & 336.5 & 323.5 & 330.9  \\
SN+RG+SDSS     & 325.1& 332.5& 325.5 & 336.5 & 324.7 & 332.1  \\
SN+RG+SDSS+CMBR& 326.5& 333.9& 326.9 & 337.9 & 325.4 & 332.8  \\
\hline
\end{tabular}
\label{tab:5}
\end{table*}


\small
\begin{thebibliography}{99}

\bibitem{Riess:2004nr}
A. Riess, et~al., Astrophys. J. 607 (2004) 665.

\bibitem{Astier:2005}
P. Astier, et~al., Astron. Astrophys. 447 (2006) 31.

\bibitem{Weinberg89}
S. Weinberg Rev. Mod. Phys. 61 (1989) 1.

\bibitem{Daly04}
R. Daly, S.G. Djorgovski,  Astrophys. J. 612 (2004) 652.

\bibitem{Casimir48}
H. Casimir, Proc. K. Ned. Akad. Wet. 51, (1948) 793.

\bibitem{Padmanabhan03}
T. Padmanabhan, Phys. Rep. 380 (2003) 335.

\bibitem{Carrol01}
S.M. Carrol, Living Reviews in Relativity 4 (2001) 1.

\bibitem{Nesterenko05}
V. Nesterenko, G. Lambiase, G. Scarpetta, Rev. Nuovo Cim. 27 (2004) 1.

\bibitem{Bordag01}
M. Bordag, V. Mohideen,  V. Mostepanenko, Phys. Rep. 353 (2005) 1.

\bibitem{McInnes06}
B. McInnes, (2006) hep-th/0607074.

\bibitem{Fischback98}
E. Fischbach, C. Talmadge, "The search for Non-Newtonian Gravity"
Springer-Verlag, New York, 1998.

\bibitem{Bressi02}
G.Bressi et al., Phys. Rev. Lett. 88 (2002) 4.

\bibitem{Ford76}
L.H. Ford, Phys. Rev. D 14 (1976) 3304.

\bibitem{Zeldovich84}
Y.B. Zeldovich, A.A. Starobinsky, Sov. Astron. Lett. 10 (1984) 135.

\bibitem{Herdeiro06}
C.A.R. Herdeiro,  M. Sampaio,  Clas. Quantum Grav. 23 (2006) 473.

\bibitem{Szydlowski87}
M. Szydlowski,  J. Szczesny,  M. Biesiada, Clas.Quantum Grav. 4 (1987) 1731.

\bibitem{Lachieze-Rey95}
M. Lachieze-Rey,  J.P. Luminet,  Phys. Rep. C 254 (1995) 135.

\bibitem{Mazur04}
P.O. Mazur, A. Mottola,  (2004) gr-qc/0405111.
 
\bibitem{Mazur06}
I. Antoniadis, P.O. Mazur, A. Mottola,  (2006) gr-qc/0612068.

\bibitem{Ishak05}
M. Ishak,  (2005) astro-ph/0504416.

\bibitem{Bean}
R. Bean,  S. Carroll,  M. Trodden, (2005) astro-ph/0510059.

\bibitem{Volovik06}
G.E. Volovik, (2006) gr-qc/0604062.

\bibitem{Altaie03}
M.B. Altaie, M.R. Setare, Phys. Rev. D 67 (2003) 044018

\bibitem{Saharian06}
A.A. Saharian, M.R. Setare, Phys. Lett. B 637 (2006) 5

\bibitem{Mahajan06}
G. Mahajan, S. Sarkar, T. Padmanabhan, (2006) astro-ph/0604265

\bibitem{Szydlowski88a}
M. Szydlowski,  J. Szczesny,  Phys. Rev D 38 (1988) 3625.

\bibitem{Szydlowski88b}
M. Szydlowski,  J. Szczesny,  T. Stawicki, Clas.Quantum Grav. 5 (1988) 1097.

\bibitem{Szydlowski89}
M. Szydlowski,  Acta Physica Polonica B 20 (1989) 671.

\bibitem{Vandersloot05}
K. Vandersloot, A. Ashtekar, P. Singh,
{\em Phenomenological implications of discreteness in loop quantum cosmology},
talk given at Loops'05, Potsdam, 10-14 October 2005; \\
http://loops05.aei.mpg.de/index\_files/abstract\_vandersloot.html

\bibitem{Singh05}
P. Singh, K. Vandersloot, Phys. Rev. D 72 (2005) 084004.

\bibitem{Hossain05}
X. Hossain, (2005) gr-qc/0504125

\bibitem{Randall99a}
L. Randal,  R. Sundrum, Phys. Rev. Lett. 83 (1999) 3370.

\bibitem{Randall99b}
L. Randal,  R. Sundrum, Phys. Rev. Lett. 83 (1999) 4690.
 
\bibitem{Arkani98}
N. Arkani-Hamed, S. Dimopoulos, G. Dvali, Phys. Lett. B 429 (1998) 263.

\bibitem{Maartens00}
R. Maartens, Phys. Rev. D. 62 (2000) 084023.

\bibitem{Vishwakarma03}
R.G. Vishwakarma, P. Singh, Clas. Quantum Grav. 20 (2003) 2033.

\bibitem{Godlowski04}
W. Godlowski, M. Szydlowski,  Gen. Rel. Grav. 36 (2004) GRG, 767.

\bibitem{Ichiki03}
K. Ichiki, et~al., Phys. Rev. D 68 (2003) 083518.
 
\bibitem{Ichiki02}
K. Ichiki, et~al., Phys. Rev. D 66 (2002) 043521.

\bibitem{Senovilla98}
J. Senovilla, F. Carlos, F. Souperta, P. Szekeres, Gen. Rel. Grav. 30 (1988) 389.

\bibitem{Heckmann59}
O. Heckmann, E. Sch{\"u}cking, Handbuch der Physik, vol. LIII,
ed. S. Fl{\"u}gge Springer-Verlag Berlin, 1959, p.489.

\bibitem{Li98}
L.-X. Li, Gen. Relat. Grav. 30 (1998) 497.

\bibitem{Godlowski03a}
W. Godlowski, M. Szydlowski, P.Flin, M.Biernacka, Gen. Relat. Grav. 35 (2003) 907.

\bibitem{Godlowski05}
W. Godlowski, M. Szydlowski, P. Flin, Gen. Rel. Grav. 37 (2005) 615.

\bibitem{Aryal06}
B. Aryal, W. Saurer, Mont. Not. Roy. Ast. Soc. 366 (2006) 438.

\bibitem{Vishwakarma04}
R.G. Vishwakarma,  (2004) astro-ph/0404371.

\bibitem{Godlowski03b}
W. Godlowski, M. Szydlowski, Gen. Rel. Grav. 35 (2003) 2171.

\bibitem{Riess98}
A. Riess, et~al., Astron. J. 116 (1998) 1009.
 
\bibitem{Daly03}
R. Daly, S.G. Djorgovski,  Astrophys. J. 597 (2003) 9.
 
\bibitem{Zhu04}
Z.H. Zhu,  M.K. Fujimoto, Astrophys. J. 603 (2004) 365.

\bibitem{Puetzfeld05}
M. Puetzfeld, M. Pohl,  Z.H. Zhu, Astrophys.J. 619 (2005) 657.

\bibitem{Godlowski06}
W. Godlowski, M. Szydlowski, Phys. Lett. B 642 (2006) 13

\bibitem{Eisenstein:2005}
D. Eisenstein, et~al.  Astrophys. J. 633 (2005) 560.

\bibitem{Wang04}
Y. Wang, M. Tegmark, Phys. Rev. Lett. 92 (2004) 241302.

\bibitem{Cardone05}
V.F. Cardone, C. Tortora, A. Troisi, S. Capozziello, Phys. Rev. D 73 (2006) 043808.

\bibitem{Peebles:2002gy}
Peebles P.~J.~E. and Ratra B., 2003 {\sl Rev. Mod. Phys.}~{\bf 75} 559

\bibitem{BA2004}
Burnham and Anderson 2002 {\sl Model Selection and Multimodel Interference:
A Practical Information-Theoretic Approach} 2nd ed. Springer Verlag, New York

\bibitem{Liddle:2004nh}
Liddle A.~R., 2004 {\sl Mon. Not. Roy. Astron. Soc.}~{\bf 351} L49.
 
\bibitem{Parkinson:2005}
Parkinson D., Tsujikawa S., Basset B. and Amendola L., 2005
 {\sl Phys. Rev.}~{\bf D71} 063524.

\bibitem{Akaike:1974}
Akaike H., 1974 {\sl IEEE Trans. Auto. Control}~{\bf 19} 716.

\bibitem{Schwarz:1978}
Schwarz G., 1978 {\sl Annals of Statistics}~{\bf 6} 461.

\bibitem{Sakamoto86}
Sakamoto Y., Ishiguro M., Katigawa G., 1986 {\sl Akaike information criterion
statistics.} Kluwer Academic Publishing, Dordrecht.

\bibitem{Jeffreys:1961}
Jeffreys H., 1961 {\it Theory of Probability}, 3rd ed., Oxford University
Press, Oxford

\bibitem{Mukherjee:1998wp}
Mukherjee S., Feigelson E.~D., Babu G.~J., Murtagh F., Fraley C. and Raftery A.,
1998 {\sl Astrophys. J.}~{\bf 508} 314.

\bibitem{Godlowski05a}
W. Godlowski, M. Szydlowski, Phys. Lett. B 623 (2005) 10.

\bibitem{Szydlowski06a}
M. Szydlowski, W. Godlowski,  Phys. Lett. B 633 (2006) 427.

\bibitem{Szydlowski06b}
M. Szydlowski, W. Godlowski,  Phys. Lett. B 639 (2006) 5.

\bibitem{Szydlowski06c}
M. Szydlowski, A. Kurek, (2006) gr-qc/0608098.

\bibitem{Szydlowski06d}
M. Szydlowski, A. Kurek, A. Krawiec, Phys.Lett. B 642 (2006) 171.

\bibitem{Molina99}
C. Molina-Paris, M. Visser, Phys. Lett. B 455 (1999) 90.

\bibitem{Tippett04}
B.K. Tippett, K. Lake, (2004) gr-qc/0409088.

\bibitem{Szydlowski05}
M. Szydlowski, W. Godlowski, A. Krawiec., J. Golbiak, Phys. Rev. D. 72 (2005) 063504.

\bibitem{Spergel:2003}
D.N. Spergel, et~al., Astrophys. J. Suppl. 148 (2003) 175.

\bibitem{Spergel06}
D.N. Spergel, et~al., (2006) astro-ph/0603449.

\bibitem{Riess06}
A. Riess, et~al., (2006) astro-ph/06111572.

\end{thebibliography}
\end{document}